\documentclass[doublecol]{epl2}

\usepackage{amsmath, color}

\title{Coexistence of dilute and densely packed domains \\of ligand-receptor bonds
in membrane adhesion}

\author{Daniel Schmidt\inst{1} \and Timo Bihr\inst{1} \and Udo Seifert\inst{1} \and Ana-Sun\v{c}ana Smith\inst{2}}
\shortauthor{D.~Schmidt, T. Bihr, U. Seifert and A.-S. Smith}
\institute{
  \inst{1} II. Institut f\"ur Theoretische Physik, Universit\"at Stuttgart, Germany\\
  \inst{2} Institut f\"ur Theoretische Physik and Excellence Cluster: Engineering of Advanced Materials, Universit\"at Erlangen-N\"urnberg, Germany
}

\abstract{We analyze the stability of micro-domains of ligand-receptor bonds that mediate the adhesion of biological model membranes. After evaluating the effects of membrane fluctuations on the binding affinity of a single bond, we characterize the organization of bonds within the domains by theoretical means. In a large range of parameters, we find the commonly suggested dense packing to be separated by a free energy barrier from a regime in which bonds are sparsely distributed. If bonds are mobile, a coexistence of the two regimes should emerge, which agrees with recent experimental observations.}

\begin{document}

\maketitle

The key step in the recognition process of living cells is the establishment of adhesive contacts either between opposing membranes of two cells or between the membrane of a cell and the extracellular matrix (ECM). It has been shown previously that the organization of bonds within domains has strong effects on the adhesion of cells and the consequent active response \cite{Geiger}. Most insightful were the experiments with cells binding to substrates containing ligands organized on a hexagonal lattice of a characteristic length between 40 and 150 nm. A  length  of 58 to 73 nm distance between bonds was shown necessary for a successful formation of domains \cite{Ranzinger}, and at distances larger then 90 nm, domains would not form \cite{DeegSpatz}.

Instead of using living cells, the so-called bottom up approach \cite{Fletcher} has been successfully used to elucidate various elements relevant to cell membranes and adhesion \cite{Smith2009}. The main protagonists of this research are giant unilamellar vesicles that are functionalized with ligands to interact with receptors immobilized on the surface \cite{Sackmann}. Depending on the density of binders on the substrate and in the vesicle,
\begin{figure}
    \onefigure[width= \linewidth]{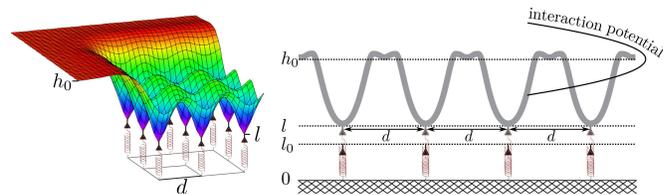}
	\caption{The system under investigation: A large patch of a bonded membrane that deforms and fluctuates in a harmonic potential. Bonds are separated by a distance $d$.}
	\label{fig1}
\end{figure}
as well as on the intrinsic binding affinity of the binding pair, either domains with densely packed bonds have been observed to grow radially from a nucleation center, or no specific adhesion was reported. In the context of the formation of these densely packed domains, valuable information on their nucleation and the growth \cite{Raudino}, equilibrium \cite{Smith2007}, cooperative effects \cite{Krobath2011, Speck2010, Reister2011}, and membrane roughness \cite{Lin2006, Krobath2007, FenzBihr2011} have been discussed over the years from both the experimental and the theoretical points of view.

With the development of experimental methods \cite{Limozin}, more detailed imaging of the distribution of bonds within the adhesion domains became possible. Consequently, large domains consisting of sparsely distributed bonds have been identified in coexistence with densely packed domains \cite{Smith2008, Smith2010}. It was reported that sparse domains may become densely packed by a gradual increase of density of ligand receptor bonds within an area of the domain of several square microns. However, some sparse domains were also found stable on time scales of the experiment (several hours) \cite{Smith2008}.

The coexistence between the sparse and dense domains driven by membrane-mediated interactions was first shown within an effective model~\cite{Bruinsma2000}. About the same time, an adhesion stabilized phase separation induced by attractive interactions of binders within the same membrane, was suggested~\cite{Komura2000}. A somewhat similar phase diagram emerged from considering the interplay of a binding bond potential and a non-specific repulsion~\cite{Andelman}, but in the absence of membrane transmitted correlations. More recently, a complex phase behavior was suggested for active binders and binders of different length coexisting within the same membrane~\cite{Weikl2009, Krobath2011}. Here we show, that the stability of sparse domains and their coexistence with dense domains emerges from basic principles, at low membrane tension. Thereby, the cost of deforming the membrane in an effective non-specific potential is balanced with the energy gain associated with the formation of bonds.

\section{The model}
We consider bonds placed on a regular square (sq) or central hexagonal (ch) lattice. Thereby, we assume that the lateral size of the membrane adhesion domain is significantly larger than the size of a lattice unit cell. We model the total free energy of a domain (containing $N$ bonds) and investigate its dependence on the distance~$d$ between the bonds.

The membrane deformation energy (in units of $k_BT$, where $k_B$ is the Boltzmann constant and $T$ the temperature) is described by the Helfrich Hamiltonian\cite{Bruinsma2000, Raedler, Andelman}
\begin{align}
 & \mathcal{H}_0 = \\
 & \int \limits_A \mathrm d{\bf r} \left [ \frac{\kappa}{2} \left (\nabla^2 h({\bf r})\right )^2 + \frac{\sigma}{2} (\nabla h({\bf r}))^2 + \frac{\gamma}{2} \left (h({\bf r})-h_0 \right )^2 \right]. \nonumber
\end{align}
Thereby, the Monge parametrization is used to represent the membrane of a bending stiffness $\kappa$ as a surface of projected area $A$ placed above the substrate at the height $h({\bf r})$, ${\bf r}=(x, y)$ being the in-plane position vector.  The first term is the bending energy, whereas the second term accounts for the tension $\sigma $ in the membrane. In a simplistic manner, the last term models the generic membrane-substrate interaction potential with the harmonic potential of a strength $\gamma$ with a minimum at the height $h_0$. In the context of mimetic systems, this interaction potential encompasses for a number of contributions such as the van-der-Waals attraction, or the steric repulsion emerging from both repeller and membrane shape fluctuations \cite{Raedler}.  In the case of cells, numerous other factors associated with actin (de)polymerization, active forces, the glycocalix and the ECM, may all contribute to this potential depending on the cell type and the treatment of the substrate.

The receptors are modeled as thermalized harmonic springs of rest length $l_0$ and spring constant $\lambda$, fixed for all bonds on the lattice. When the receptor is relatively stiff such as a bulky protein, it is modeled with a very large spring constant $(\lambda\to \infty)$. If the receptor is a soft polymer, deforming to form a bond, $\lambda$ is set finite.

Ligands and receptors interact through a square-well potential~\cite{Andelman,Breidenich}, of a very short range $\alpha$ and depth $\epsilon_b$ ($1-35~k_\textrm{B}T$), the latter associated with the intrinsic binding affinity. Thus, the total Hamiltonian $\mathcal{H}$ of a domain with $N$ bonds situated at position ${\bf r}_b^j$ reads
\begin{equation}
  \mathcal{H} = \mathcal{H}_0 + \frac \lambda2 \sum_{j=1}^N (l({\bf r}_b^j)-l_0)^2  + N\epsilon_b\equiv N\mathcal{H}_d+ N\epsilon_b.
 \label{eq_membraneenergy}
\end{equation}
Here, $l({\bf r}_b^j)$ is the extension of the $j$-th spring. When $\alpha\rightarrow 0$, obviously $h({\bf r}_b^j)\rightarrow l({\bf r}_b^j)$. Furthermore, $\mathcal{H}_d$ denotes the deformation energy per bond stored in the membrane and all receptors. The last term is the binding enthalpy.

\section{Free energy} The stability of the domain is determined from the difference $\Delta \mathcal F^N$ between the free energy of the domain with $N$ formed bonds $\mathcal F_b^N$, and the free energy of the reference state in which receptors and the membrane fluctuate freely $\mathcal F_{ub}^N$. Both $\mathcal F_b^N$ and $\mathcal F_{ub}^N$ are calculated from the partition function $\mathcal Z$, comprising all possible conformations of $N$ receptors and the membrane. Thereby, the partition functions of the reference and the bound state ($\mathcal Z_{ub}$ and $\mathcal Z_{b}$), and hence $\mathcal F_{ub}^N\equiv \ln\mathcal Z_{ub}$ and $\mathcal F_{b}^N\equiv \ln\mathcal Z_{b}$, are associated with the conformations in which the membrane is outside or within the range $\alpha$ of the square potential at the position of all receptors, respectively. The domain is stable if $\Delta \mathcal F^N=\mathcal F_b^N-\mathcal F_{ub}^N<0$.

Here we have clearly omitted the change in the mixing entropy of binders that could, in principle affect the results. However, we assume the adhesion to be mediated by large domains of bonds. Such domains typically form when the concentration of the mobile ligands is significantly larger than the concentration of the immobilized receptors~\cite{Smith2007}.  In the context of the free energy of the system, in this regime the mixing entropy will provide a constant, proportional to the number of bonds and to the chemical potential of the free ligands, but independent on the distance between bonds. Consequently, in this regime, the mixing entropy will act to re-scale the effective binding affinity of a single bond~\cite{Bruinsma2000}, without affecting the general phase behavior of the domains.

\subsection{The reference state for $N$ receptors}

The partition function for the free membrane and $N$ unbound fluctuating receptors is, up to the normalization,
\begin{align}
\mathcal Z_{ub} = & \prod_{j=1}^N \left (\int \mathrm dl({\bf r}_b^j) \exp \left [ -\frac{\lambda}{2} (l({\bf r}_b^j)-l_0)^2 \right ]  \right ) \nonumber \\
& \times  \left ( \int \mathcal{D} \left [h^\prime ({\bf r}) \right ] {\rm e}^{- \mathcal H_0 [h^\prime ({\bf r})]} \right ) \equiv \mathcal C \left(\frac{2\pi}{\lambda} \right)^{N/2}
\end{align}
where $\mathcal C$ is denoting the result of the functional integral $\mathcal{D} \left [h^\prime_{\bf q} \right ]$.

In this state, the membrane is free, on average flat, and positioned in the minimum of the nonspecific potential  $\langle h({\bf r})\rangle=h_0$. The fluctuation amplitude is that of a the membrane under tension~\cite{LipowskyBook}
 \begin{equation}
\label{eq_flucampl}
 \Sigma_0^2 \equiv \frac{1}{A} \sum_{\bf q} \frac{1}{\kappa q^4 + \sigma q^2 + \gamma}=\frac{\arctan(\sqrt{4\kappa \gamma-\sigma^2}/\sigma)}{2\pi\sqrt{4\kappa \gamma-\sigma^2}}
\end{equation}
where the sum runs over all possible wave vectors ${\bf q}$ with $q\equiv|{\bf q}|$, set by the system size. The spatial correlation function ~\cite{Bruinsma94, LipowskyBook} is simply given by
 \begin{equation}
\label{eq_corr}
 G({\bf r}_0 - {\bf r})\equiv \frac{1}{A} \sum_{\bf q} \frac{\cos\left({\bf q}\cdot({\bf r}_0 - {\bf r}) \right)}{\kappa q^4  + \sigma q^2 + \gamma},
\end{equation}
where ${\bf r}_0$ and ${\bf r}$ are arbitrary positions on the membrane. In the tensionless limit, $\Sigma_0^2=1/(8\sqrt{\kappa \gamma})$ and $G(r)\approx -4\pi^{-1}\xi_{\bot}^2\mathrm{kei} (r/\xi_{\|})$, with $kei$ signifying the kelvin function. Thereby, $\xi_{\bot}^2 \equiv 1/(8\sqrt{\kappa \gamma})$ and $\xi_{\|} \equiv \sqrt[4]{\kappa / \gamma}$ are the vertical roughness and the lateral correlation length of a tensionless unbound membrane, respectively~\cite{Raedler, Bruinsma94, LipowskyBook}, setting the length and other scales of the system.

\subsection{An isolated bond - $N=1$}
The shape of the membrane bound to the substrate by one bond is typified by the Kelvin function \cite{Bruinsma94}. The associated membrane deformation energy $\mathcal{U}(\sigma)=(h_0-h({\bf r}_b))^2 /(2\Sigma_0^2)$ is quadratic with respect to the offset from the minimum of the nonspecific potential . When $\lambda\to \infty$ the entire deformation is stored in the membrane. If, furthermore, $\sigma=0$, one finds $\mathcal{U}_0=(h_0-l_0)^2/2\xi_{\bot}^2$, as previously determined \cite{Bruinsma2000}.

\begin{figure}[t]
 \begin{center}
  \onefigure[width=1 \linewidth]{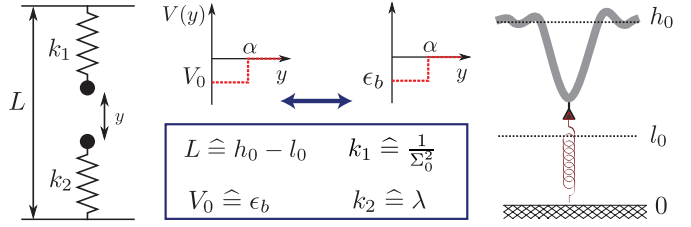}
  \caption{Mapping of the model for two fluctuating interacting springs to the membrane-bond model.}
 \label{onebond}
 \end{center}
\end{figure}

Because both, the membrane and the receptor deformations are quadratic with the elongation, we map the problem of forming one bond to a problem of two one-dimensional thermalized springs of stiffness $k_1$ and $k_2$ (Fig. \ref{onebond}). The springs are said to interact if their relative distance falls within a square-well potential of a (short) range~$\alpha$ and depth $V_0$. The the free energy difference between the bound and the unbound state 
\begin{equation}
 \Delta \mathcal F^{sp} = \frac{1}{2}\frac{k_1k_2 L^2}{k_1+k_2} - V_0 + \frac{1}{2} \ln \left [ \frac{2\pi (k_1+k_2)}{k_1k_2} \frac{1}{\alpha^2} \right ]
\end{equation}
is calculated as described previously, by subdividing the configurational space of the receptor and the membrane. The first term on the right side is identified with the deformation energy of the two springs, characterized by a reduced spring constant of the coupled system consisting of two springs in series $k_1k_2/(k_1+k_2)$ elongated to meet the system size $L$. The second and the third terms are the enthalpy gain and the entropy loss due to the formation of a bond. Thereby, the last contribution only affects the depth of the potential.

By analogy (Fig.~\ref{onebond}), for a membrane binding to a single receptor (indicated by the superscript $1$)
\begin{equation}
 \Delta \mathcal F^{1} = \frac12 \frac{(h_0-l_0)^2}{\Sigma_0^2+1/\lambda} - \bar{\epsilon_b} \equiv \mathcal{H}_{d}^{1}-\bar{\epsilon_b}.
\label{eq_freeenergydifference1bond}
\end{equation}

Thereby, $\mathcal{H}_{d}^{1}$ is the total deformation energy associated with an isolated bond and sets the energy scale of the problem. For $\sigma=0$ and $\lambda \rightarrow \infty$, $\mathcal{H}_{d}^{1}=\mathcal{U}_0$. Furthermore,  $\bar{\epsilon_b}$ is the effective binding affinity
\begin{equation}
 \bar{\epsilon_b} \equiv \epsilon_b - \frac 12 \ln \left[ \frac{2\pi}{\alpha ^2} \left ( \frac{1}{\lambda} + \Sigma_0^2 \right ) \right],
\label{eq_effbindingaffinity}
\end{equation}
which is the contribution of a single bond to the free energy. In essence, it is the intrinsic binding affinity decreased by the entropic cost related to the change in the fluctuations of the membrane and the receptor upon binding. At room temperature and typical parameters this cost amounts to several $k_BT$.
If mixing entropy of the binders is considered in the regime of large domains, the chemical potential of free ligands would need to be subtracted on the right hand side of the eq.~(\ref{eq_effbindingaffinity}).

\subsection{ A domain with $N$ bonds}

The free energy of an arbitrary bond configuration of $N$ bonds emerges from the partition function for the bound system $\mathcal Z_b$
\begin{equation}
\mathcal Z_b = \int \mathcal{D} \left [ h^\prime({\bf r}) \right ] \prod_{j}^{N} \int\limits_{h^\prime({\bf r}_b^j)-\alpha}^{h^\prime({\bf r}_b^j)} \mathrm dl({\bf r}_b^j) {\rm e}^{-\mathcal H[h^\prime({\bf r}), \{l({\bf r}_b^j)\}]},
\label{eq_boundpartionfunction}
\end{equation}
that accounts for all conformations in which all receptors and the membrane are simultaneously within the bond potential range $\alpha$. With the Hamiltonian from eq. (\ref{eq_membraneenergy}), one gets
\begin{align}
\mathcal Z_b = \frac{\mathcal C \alpha^N \mathrm{exp}{\left[-\left (\mathcal{H}_{d}^{1} \sum_{i,j=1}^N M^{-1}_{ij}-N\epsilon_b\right)\right]}}{\sqrt{(1 + \lambda \Sigma_0^2)^N \det M}} ,
\end{align}
with $M_{ij} \equiv (\delta_{ij} + \lambda G({\bf r}_b^i - {\bf r}_b^j))/(1+\lambda \Sigma_0^2)$ accounting for the membrane-coupled deformations of bound receptors on positions ${\bf r}_b^i$ and ${\bf r}_b^j$. The free energy difference becomes
\begin{equation}
 \Delta \mathcal F^N = \mathcal{H}_{d}^{1} \sum_{i,j=1}^N M^{-1}_{ij}  - N \bar{\epsilon_b} + \frac 12 \ln \left ( \det M \right ).
\label{eq_freeenergydifference}
\end{equation}
Similarly to eq. (\ref{eq_freeenergydifference1bond}), the first term in eq. (\ref{eq_freeenergydifference}) is the total deformation energy of the receptors and the membrane. The second term is proportional to the effective binding affinity of a single bond and to the total number of bonds within the domain. The last term calculates the fluctuation-induced interactions between the bonds, which interestingly, can be fully decoupled from other contributions.  At large bond separations, $d\gg \xi_{\|}$, $\mathcal{H}_{d}\approx\mathcal{H}_{d}^{1}$ and $\Delta \mathcal{F}^N/N \approx \Delta\mathcal{F}^{1}$.

\subsection{Phase diagram} 
\begin{figure}[t]
 \begin{center}
  \onefigure[width=1 \linewidth]{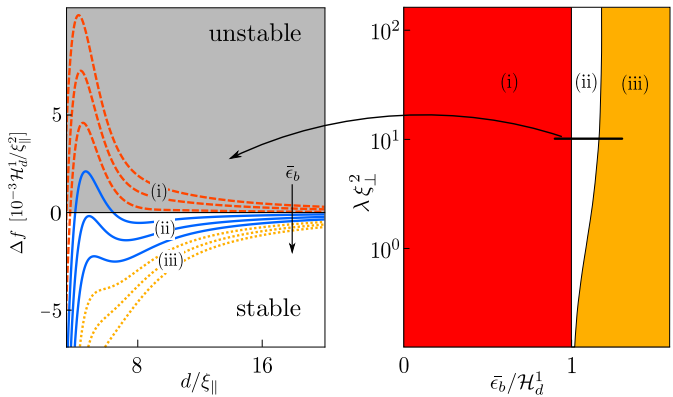}
  \caption{Free energy density as a function of the distance between the bonds for the ch lattice (left), and the respective phase diagram (right). All variables are rendered dimensionless. $\sigma=0$. For a detailed discussion, see the main text.}
 \label{fig5}
 \end{center}
\end{figure}

To compare domains with the same number of bonds that due to packing, cover different area, we analyze the free energy density $\Delta f \equiv \Delta\mathcal{F}^N/A$ that if negative, signifies stable domains (fig. \ref{fig5}). Three distinct regimes are apparent even though a global minimum is found in all cases. This minimum, however, appears at such small $d$ that it could be rendered inaccessible by the finite size of the binders.

In the limit when $d\to \infty$, the $\Delta f \to 0$ limit is approached either from positive values (fig. \ref{fig5}, red-dashed lines) or negative values (blue and yellow-dotted lines), signifying unstable and stable binding, respectively. In the former case (regime ($i$) in the phase diagram), by a maximum at intermediate distances precedes either stable (regime ($ii$)) or unstable domains at large distances between the bonds (regime ($i$)), in which case the domains are stable only close to the boundary minimum. The phase transition line separating the regimes $(i)$ and $(ii)$ is found in the limit of $d\to \infty$ as $\bar{\epsilon_b} = \mathcal{H}_{d}^{1}$

Following the maximum, $\Delta f$ curves of the region ($ii$) possess a shallow secondary minimum that is always negative, with locally stable domains. The maximum, on the other hand, penetrates toward positive values of $\Delta f$ for smaller values of $\bar{\epsilon_b}$, making the domains unstable at intermediate bond separations.  At larger values of $\bar{\epsilon_b}$,  the maximum may remain with fully negative free energies, hence stable domains can take place for any distances between bonds. This is also true for the regime $(iii)$ in which $\Delta f$ remains negative, despite its monotonous increase. The border between the regimes ($ii$) and ($iii$) is given by the (dis)appearance of a root in the second derivative of the free energy density with respect to the bond distance, and has to be determined numerically.

This phase diagram should be applicable to the situation in which the distance between the bonds is predefined, such as when one of the binder type is immobilized on the substrate \cite{DeegSpatz, Sackmann}. The domains should be observable for any distance in which the total free energy density is smaller than zero. Consequently, the commonly observed densely packed domains \cite{Smith2010} can be found within our model at low free energies. However, when both receptors and ligands are mobile, the distance between the bonds within the domain becomes a free parameter. In this case, domains should be found only at the distances at which the minima in the total energy density appear. More specifically, apart from the densely packed agglomerates, domains associated with the minimum in the region ($ii$) at intermediate distances between the bonds, should be seen. Indeed, coexistence between densely packed domains and domains with a sparse distribution of bonds has been observed recently in experiments with mobile ligand-receptor pairs \cite{Fenz2011, Smith2008}.

\section{Optimum deformation} The understanding of the above phase diagram evolves from the analysis of the mean membrane shape and the fluctuations amplitude. They emerge as moments of the height probability distribution $p(h({\bf r}))$ of the membrane within the domain with a fixed bond configuration. The latter is a functional integral over all appropriately weighed realizations of the membrane profile
\begin{align}
 p\left(h({\bf r})\right) & \sim \int \mathcal{D}[h^\prime({\bf r})] \exp(-\mathcal H[h^\prime({\bf r})]) \delta(h^\prime({\bf r}) - h({\bf r})) \nonumber \\
 & \sim  \exp \left( -\frac 12 \frac{(h({\bf r}) - \langle h({\bf r}) \rangle)^2}{ \Sigma^2({\bf r})} \right ).
 \label{eq_partitionfunction}
\end{align}
Because of the quadratic form of eq.~(\ref{eq_membraneenergy}), $p\left(h({\bf r})\right)$ is a Gaussian distribution with the expectation value giving the equilibrium shape
\begin{equation}
\label{shape2}
 \langle h({\bf r}) \rangle \equiv h_0 - \frac{   ( h_0-l_0 ) }{\Sigma_0^2 /  \Sigma^2({\bf r})}\sum_{ij}G({\bf r}_b^i - {\bf r}) L^{-1}_{ij},
\end{equation}
 and variance $\Sigma^2({\bf r})$
\begin{equation}
  \Sigma^2({\bf r})\equiv \frac{ \Sigma_0^4}{ \Sigma_0^2+ \sum_{ij} G({\bf r}_b^i - {\bf r}) L^{-1}_{ij} G({\bf r}_b^j - {\bf r})},
\end{equation}
being the fluctuation amplitude. Thereby
\begin{equation}
 L_{ij} \equiv \frac{\delta_{i,j}}{\lambda}+ G({\bf r}_b^i - {\bf r}_b^j) - \frac{G({\bf r}_b^i - {\bf r})G({\bf r}_b^j - {\bf r})}{\Sigma_0^2}.
\end{equation}
By setting $N=0$ and $N=1$ one recovers the results presented in previous sections.

The equilibrium shape can be also determined by direct minimization of $\mathcal{H}$ from eq. (\ref{eq_membraneenergy}), with constrained extension of bonds, and periodic boundary conditions. Consequently, the effect of the lattice is explicit, 
 and one obtains
\begin{equation}
 \langle h ({\bf r}) \rangle = h_0 - \frac{g({\bf r})}{\phi+1/\lambda} (h_0-l_0).
 \label{eq_shape}
\end{equation}
Here $\phi \equiv a^{-1} \sum_{\bf q} (\kappa q^4 + \sigma q^2 + \gamma)^{-1}$ and $ g({\bf r}) \equiv a^{-1} \sum_{\bf q} \cos({\bf qr})(\kappa q^4  + \sigma q^2 + \gamma)^{-1}$, with $a$ being the area of a unit cell. For the squared lattice, the sums run over wave vectors ${\bf q} \equiv (q_1, q_2)= (2\pi z_{1}/ d, 2\pi z_{2}/ d)$, with $z_{1}, z_{2}$ being integers, while for the hexagonal lattice, $q_1=2\pi (2z_1+z_2) /(\sqrt{3}d)$ and $q_2= 2 \pi z_2/ d$. 

Combining the total Hamiltonian, eq. (\ref{eq_membraneenergy}), and the equilibrium shape of the membrane, eq. (\ref{eq_shape}), results in the total membrane and spring deformation energy per bond
\begin{equation}
 \mathcal{H}_d [\langle h ({\bf r}) \rangle] = \frac{(h_0-l_0)^2}{2 \left [ \phi + 1/\lambda \right ]}.
 \label{eq_totalmembraneenergy}
\end{equation}
Thereby, all bonds have the same extension and the membrane is at $h({\bf r}_b^i)=h_b$, for all ${\bf r}_b^i$. This result is consistent with the first term in eq.(\ref{eq_freeenergydifference}) as well as with the 1D energy profile found for a membrane deformed by two infinite cylinders \cite{Weikl2003}, and the potential calculated for the interaction between two bonds \cite{Bruinsma2000}.

\begin{figure}[tbp]
 \begin{center}
  \onefigure[width=1 \linewidth]{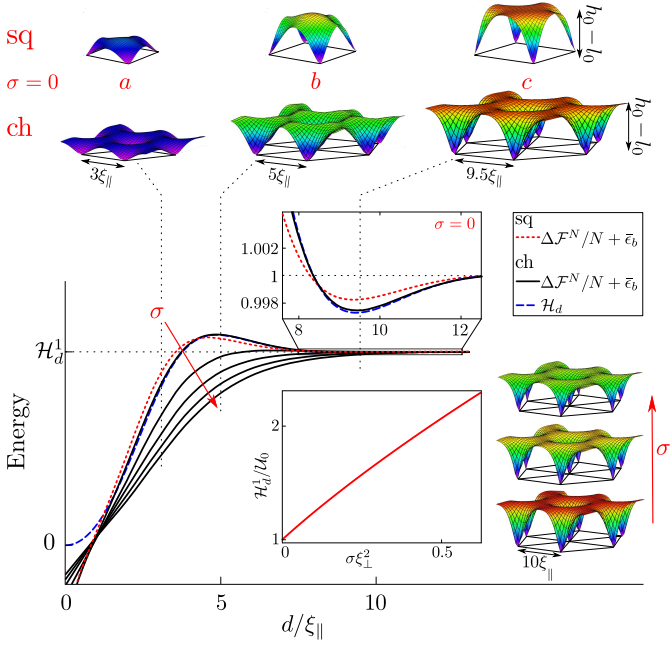}
  \caption{For $\sigma=0$, stiff bonds, and $\mathcal{H}_d^1(\sigma=0)=5.6~k_\textrm{B}T$, shapes and the free energy per bond $\Delta\mathcal{F}^N/N+ \bar{\epsilon_b}$ for the ch and the sq lattice (black and red dotted line, respectively) are displayed. The deformation energy per bond $\mathcal{H}_d$ on the ch lattice is presented with the blue dashed line. Upper inset highlights the minimum at $d=9.5 \xi_\parallel$. For tensions $\sigma=(0.125,0.25,0.375,0.5)\xi_\perp^{-2}$, shapes (ch lattice, $d=10 \xi_\parallel$) and the free energy are shown. All curves are scaled by $\mathcal{H}_d^1$ that depends on the tension as shown in the lower inset.}
 \label{fig2}
 \end{center}
\end{figure}

\begin{figure}[t]
 \begin{center}
  \onefigure[width=1 \linewidth]{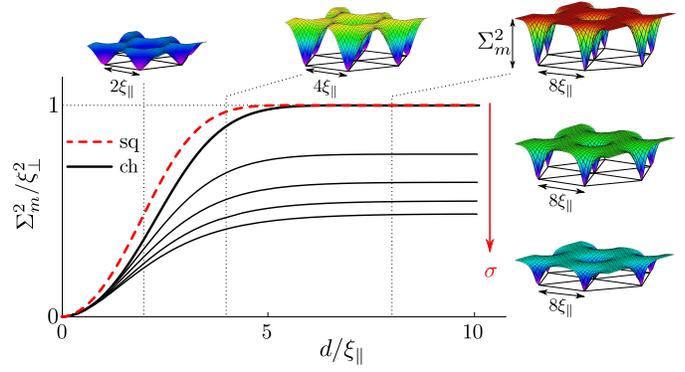}
  \caption{ Maximal fluctuation amplitude for the ch and sq lattice (black and red-dashed lines, respectively). Fluctuation maps at $\sigma=0$ (horizontal array), and finite tensions at $d=8\xi_\parallel$ (vertical array) are also shown. Parameters as in Fig. 4.}
 \label{fig3}
 \end{center}
\end{figure}

It is instructive to first analyze the case of stiff bonds when $\lambda(h({\bf r_b})-l_0)^2\to 0$, and $ \mathcal{H}_d$, eq. (\ref{eq_totalmembraneenergy}), depends only on the lattice type and $d$ (fig. \ref{fig2}). The optimum deformation energy has a shallow minimum at intermediate distances between the bonds, causing the minimum in the free energy density. This minimum can be understood by analyzing the shape of the membrane. Namely, when the bonds are far apart, they act as isolated bonds producing a local deformation in which the membrane, prior to flattening into the minimum of the nonspecific potential at $h_0$, overshoots $h_0$ (fig. \ref{fig2}). As the bonds come closer, the overshoots become shared by neighboring bonds, decreasing the overall cost in bending. When the overshoots fully overlap, the shallow minimum appears in $\Delta f$ (shapes (\textit{c}) in  fig. \ref{fig2}). Bringing the bonds even closer, again increases the bending energy providing an energy barrier (shapes (\textit{b})). When the overshoots disappear the membrane starts to flatten between the bonds. Consequently, the energy slides towards a boundary minimum $\mathcal{H}_d\to 0$, as $d \to 0$. With the increasing tension (at constant $\gamma$) the overshoots in the shape become less pronounced, but the cost for deforming the membrane  rise (fig. \ref{fig2}). The secondary minimum becomes shallower and appears at larger $d$, shrinking the parameter space for the phase (\textit{ii}).

In the limit of $d\to0$, flattening of the membrane caused by the large density of bonds makes both the deformation energy and fluctuations insensitive to the tension (fig. \ref{fig3}). If, on the other hand, $d \to \infty$, bonds on any lattice become independent from one another and the maximal fluctuation amplitude $\Sigma_\mathrm{m}^2$ of the domain fluctuation profile $\Sigma({\bf r})^2$, takes the limit $ \Sigma_\mathrm{m}^2\to\Sigma_0^2$.

\begin{figure}[t]
 \begin{center}
  \onefigure[width=0.9 \linewidth]{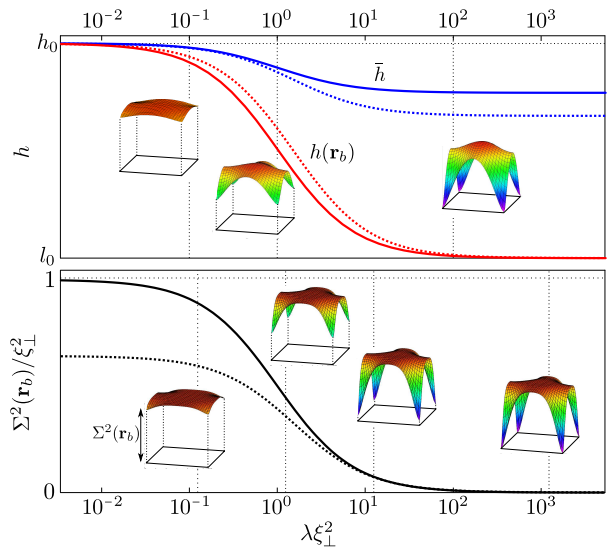}
 \caption{Top: Average height of the membrane $\bar h$ (blue) and the extension of the bonds (red) as a function of $\lambda$ and the corresponding shapes at $d=6\xi_\parallel$. Bottom: The fluctuation amplitude of a bond as a function of $\lambda$. Full and dotted lines are obtained with $\sigma=0$ and $\sigma = 1/(4\xi_\perp^2)$, respectively.}
 \label{fig4}
 \end{center}
\end{figure}

Decreasing the spring constant $\lambda$ transmits the deformation from the membrane to the receptors, again affecting the size of the region (ii) in the phase diagram. For very soft bonds ($\lambda\to 0$), the entire deformation is stored in the springs and ($h({\bf r}_b)\to h_0$). In this case, the membrane is in average flat as evidenced by the mean membrane height $\bar h$ within the domain and the bond extensions (fig. \ref{fig4}). Furthermore, in this regime, the membrane fluctuates as an unbound one, irrespectively of $d$.

\section{Conclusions}

We analyzed the properties of large adhesion domains forming between a membrane and a flat substrate, when the ligand-receptor adhesion competes with the nonspecific adhesion. While some aspects of our model have been investigated previously \cite{Andelman, Bruinsma2000}, our work provides a unifying framework within which the stability of the domains can be fully explored. Our first conclusion is that the energetics of the domains forming on different lattices including the simple hexagonal one (data not shown) is qualitatively the same. This suggests that the modeling on commonly used sq lattices \cite{LipowskyBook,Merath} will well reproduce the behavior of domains that most likely form on the ch lattice. This result emerges from the decoupling of the entropic free energy contributions associated with the correlations between bonds and contributions of individual bonds. At room temperatures, the correlation contributions seem to be small and have no qualitative effects on the phase diagram. However, on a level of individual bonds, the effect of fluctuations may be significant resulting in considerable differences between the intrinsic and the effective binding affinity. Furthermore, another important result is the evaluation of several regimes in which domains are stable. Consequently, our model provides a physical explanation for the recently observed coexistence of densely packed and sparse domains of bonds.

While we have demonstrated the power of our approach with a very simple model of receptors and of the bond potential, the quadratic nature of the membrane deformation energy allows for stability analysis of domains formed from bonds with a very wide range of potentials. Such extensions of the model may be important for quantitative comparison with measured data, which is a task that may be challenging, both from experimental and theoretical points of view. However, in order to fully comprehend the stability of adhesion domains, this comparison needs to be performed. In this light, the current work could become the necessary foundation for the understanding of the stability of domains in cellular and mimetic systems.

\section{Acknowledgements}
We thank K.~Sengupta and S. Fenz for helpful discussions. US, A-SS and TB acknowledge the support from
DFG-SE 1119/2-1.


\begin{thebibliography}{10}

\bibitem{Geiger}
Geiger B., Spatz J. P., and Bershadsky A. D.
\newblock {\em Nat. Rev. Mol. Cell Bio.} 10:21--33, 2009

\bibitem{Ranzinger}
Ranzinger J. et al.
\newblock {\em Nano Lett.} 9:4240--4245, 2009

\bibitem{DeegSpatz}
Deeg J. A. et al.
\newblock {\em Nano Lett.} 11:1469--1476, 2011

\bibitem{Fletcher}
Liu A. P. and Fletcher D. A.
\newblock {\em Nat. Rev. Mol. Cell Bio.} 10:644--650, 2009

\bibitem{Smith2009}
Smith A.-S. and Sackmann E.
\newblock {\em ChemPhysChem} 10:66--78, 2009

\bibitem{Sackmann}
Sackmann E. and Bruinsma R. F.
\newblock {\em ChemPhysChem} 3:262--269, 2002;
Lorz B. et al.
\newblock {\em Langmuir}, 23:12293--12300, 2007.

\bibitem{Raudino}
Raudino A. and Pannuzzo M.
\newblock {\em J. Chem. Phys.}, 132:045103, 2010.

\bibitem{Smith2007}
Smith A.-S. and Seifert U.
\newblock {\em Soft Matter}, 3:275--289, 2007.

\bibitem{Speck2010}
Speck T., Reister E., and Seifert U.
\newblock {\em Phys. Rev. E}, 82:021923, 2010;
Farago O.
\newblock {\em Phys. Rev. E}, 78:051919, 2008.

\bibitem{Reister2011}
Reister E. et al.
\newblock {\em New J. Phys.}, 13:025003, 2011.

\bibitem{Krobath2011}
Krobath H., Rozycki B., Lipowsky R., and Weikl T. R.
\newblock {\em PLoS One}, 6(8):e23284, 2011.

\bibitem{Krobath2007}
Krobath H., Sch\"utz G. J., Lipowsky R., and Weikl T. R.
\newblock {\em EPL}, 78:38003, 2007.

\bibitem{FenzBihr2011}
Fenz S. F. et al.
\newblock {\em Adv. Mater.}, 23:2622, 2011.

\bibitem{Lin2006}
Lin L. C.-L., Groves J. T., and Brown F. L. H.
\newblock {\em Biophys. J.}, 91:3600--3606, 2006.

\bibitem{Limozin}
Limozin L. and Sengupta K.
\newblock {\em ChemPhysChem}, 10:2752--2768, 2009;
Monzel C., Fenz S. F., Merkel R., and Sengupta K.
\newblock {\em ChemPhysChem}, 10:2828--2838, 2009.

\bibitem{Smith2010}
Smith A.-S., Fenz S. F., and Sengupta K.
\newblock {\em EPL}, 89:28003:1--6, 2010.

\bibitem{Smith2008}
Smith A.-S. et al.
\newblock {\em Proc. Natl. Acad. Sci. U. S. A.}, 105(19):6906--6911, 2008.

\bibitem{Bruinsma2000}
Bruinsma R., Behrisch A., and Sackmann E.
\newblock {\em Phys. Rev. E}, 61:4253--4267, 2000.

\bibitem{Komura2000}
Komura S. and Andelman D.
\newblock {\em Eur. Phys. J. E.}, 3:259-271, 2000.

\bibitem{Andelman}
Weikl T. R., Andelman D., Komura S., and Lipowsky R.
\newblock {\em Eur. Phys. J. B}, 8:59--66, 2002.

\bibitem{Weikl2009}
Weikl T.~R. et. al.
\newblock {\em Soft Matter}, 5:3273, 2009.

\bibitem{Raedler}
Raedler J. O., Feder T. J., Strey H. H., and Sackmann E.
\newblock {\em Phys. Rev. E}, 51(5):4523--4536, 1995.

\bibitem{Breidenich}
Breidenich M., Netz R. R., and Lipowsky R.
\newblock {\em Eur. Phys. J. E}, 5:403--411, 2001.

\bibitem{LipowskyBook}
Lipowsky R. in \newblock {\em Structure and Dynamics of Membranes}. Editors Lipowsky R. and Sackmann E., chapter 11, Elsevier, 1995.

\bibitem{Bruinsma94}
Bruinsma R., Goulian M., and Pincus P.
\newblock {\em Biophys. J.}, 67:746--750, 1994.

\bibitem{Fenz2011}
Fenz S. F., Smith A.-S., Merkel R., and Sengupta K.
\newblock {\em Soft Matter}, 7(3):952--962, 2011.

\bibitem{Weikl2003}
Weikl T. R.
\newblock {\em Eur. Phys. J. E}, 12:265--273, 2003.

\bibitem{Merath}
Merath R.-J. and Seifert U.
\newblock {\em Phys. Rev. E}, 73:010401, 2006.



\end{thebibliography}
\end{document}